\documentstyle[osa,psfig]{revtex}
\newcommand{\beq}{\begin{equation}}
\newcommand{\eeq}[1]{\label{#1} \end{equation}}
\newcommand{\beqar}{\begin{eqnarray}}
\newcommand{\eeqar}[1]{\label{#1} \end{eqnarray}}
\newcommand{\insertplot}[1]{
     \centerline{\psfig{figure={#1},height=9.0cm}}
}
\begin{document}

\begin{center}
{\large {\bf Freeze out in hydrodynamical models }} \bigskip

Cs. Anderlik,$^{1,2}$ L.P. Csernai,$^{1,2,3}$ F. Grassi,$^4$ W. Greiner,$^2$
Y.~Hama,$^4$ T.~Kodama,$^5$ Zs.I. L\'az\'ar,$^{1,2,6}$ V.K. Magas$^1$ and H.
St\"ocker$^2$ \bigskip

$^1$Section for Theoretical Physics, Department of Physics\\University of
Bergen, Allegaten 55, 5007 Bergen, Norway\\$^2$ Institut f\"ur Theoretische
Physik, Universit\"at Frankfurt\\Robert-Mayer-Str. 8-10, D-60054 Frankfurt
am Main, Germany\\$^3$ KFKI Research Institute for Particle and Nuclear
Physics\\P.O.Box 49, 1525 Budapest, Hungary\\[0.2ex]$^4$Instituto de
F\'{\i}sica, Universidade de Sao Paulo\\CP 66318, 05389-970 S\~ao Paulo-SP,
Brazil\\$^5$Instituto de F\'{\i}sica, Universidade Federal do Rio de Janeiro%
\\CP 68528, 21945-970 Rio de Janeiro-RJ, Brazil \\$^6$Department of Physics,
Babe\c{s}-Bolyai University\\Str. M. Kog\u{a}lniceanu nr. 1, 3400
Cluj-Napoca, Romania \bigskip

{\it Version R1.04 Jan. 25, 1999}\bigskip 
\end{center}

{\bf Abstract:} We study the effects of strict conservation laws and the
problem of negative contributions to final momentum distribution during the
freeze out through 3-dimensional hypersurfaces with space-like normal. We
study some suggested solutions for this problem, and demonstrate it on one
example. \bigskip

\noindent PACS: 24.10.Nz, 25.75.-q\\\noindent Keywords: freeze out; particle
spectra, conservation laws.

\section{Introduction}

\indent Fluid dynamical models, especially their simpler versions are very
popular in heavy ion physics, because they connect directly collective
macroscopic matter properties, like the Equation of State (EoS) or transport
properties, to measurables.

Particles which leave the system and reach the detectors, can be taken into
account via source (drain) terms in the 4-dimensional space-time based on
kinetic considerations, or in a more simplified way via freeze out (FO) or
final break-up schemes, where the frozen out particles are formed on a
3-dimensional hypersurface in space-time. This information is then used as
input to compute measurables such as two-particle correlation, transverse-,
longitudinal-, radial-, and cylindrical- flow, transverse momentum and
transverse mass spectra, etc.

In this paper we concentrate on freeze out. A basic standard assumption in
this case is that freeze out happens across a hypersurface as already
mentioned, so it can be pictured as a discontinuity where the kinetic
properties of the matter, such as energy density and momentum distribution
change suddenly. The hypersurface is an idealization of a layer of finite
thickness (of the order of a mean free path or collision time) where the
frozen-out particles are formed and the interactions in the matter become
negligible. The dynamics of this layer is described in different kinetic
models such as Monte Carlo models \cite{BB95,BB97} or four-volume emission
models \cite{BC82,GH95,GH96,GH97,He97}. In fact, the zero thickness limit of
such a layer is an over-idealization of kinetic freeze out in heavy ion
reactions, while it is applicable on more macroscopic scales like in
astrophysics \footnote{%
On the other hand, if kinetic freeze out coincides with a rapid phase
transition, like in the case of rapid deconfinement transition of
supercooled quark-gluon plasma, the sharp freeze out hypersurface
idealization may still be applicable even for heavy ion reactions. It is,
however, beyond the scope of this work to study the freeze out dynamics and
kinetics in this latter case.}.

Two types of hypersurfaces are distinguished: those with a space-like normal
vector, $d\sigma ^\mu d\sigma _\mu =-(d\sigma )^2$ (e.g. events happening
on a propagating 2-dimensional surface) and those with a time-like normal
vector $d\sigma ^\mu d\sigma _\mu =(d\sigma )^2$ (a common example of
which is an overall sudden change in a finite volume).

Once the freeze out surface is determined, one can compute measurables.
Landau when drafting his hydrodynamical model\cite{La53}, just evaluated the
flow velocity distribution at freeze out, and this distribution served as a
basis for all observables. This approach was used in early fluid dynamical
simulations of heavy ion collisions also \cite{SM74,AH757,AG78}. This
procedure was improved to add thermal velocities to the flow velocities at
freeze out, by Milekhin\cite{Mi58,Mi61} and later by Cooper and Frye~\cite
{CF74}. This method is widely used, however it rises at least three problems 
\cite{CLM97}.

{\em First,} in some cases before the 90's, the possible existence of
discontinuities across hypersurfaces with time-like normal vectors was not
taken into account or considered unphysical\footnote{%
Taub\cite{Ta48} discussed discontinuities across propagating hypersurfaces,
which have a space-like normal vector. If one applies Taub's formalism from
1948, to freeze out surfaces with time-like normal vectors, one gets a usual
Taub adiabat but the equation of the Rayleigh line yields imaginary values
for the particle current across the front. Thus, these hypersurfaces were
thought unphysical. However more recently, Taub's approach has been
generalized to these hypersurfaces \cite{Cs87} (see also \cite{Cs94}) while
eliminating the imaginary particle currents arising from the equation of the
Rayleigh line. Thus, it is possible to take into account conservation laws
exactly across any surface of discontinuity with relativistic flow.} \cite
{GK84,DR87,LLT92,BC93}. This point was studied recently \cite{GMQ94} so we
do not discuss it further.

{\em Second, }since the kinetic properties of the matter are different on
the two sides of the front, the explicit evaluation of conservation laws
across the freeze out surface should be taken into account which is not
always easy to implement. In some (simple) cases \cite{GS84,Si89,Bu96},
these conservation laws are enforced and discussed. For example in \cite
{GS84}, it was pointed out that the freeze out momentum distribution for
hypersurfaces with time-like normal may become locally anisotropic. In \cite
{LANL-XXX} a solution for post FO massless, baryonfree Bose gas was
presented. Here we remind the procedure that should be followed in section 
\ref{two}.

The {\em third} problem is a conceptual problem arising in the Cooper-Frye
freeze out description when we apply it to a hypersurface with space-like
normal: it is the problem of negative contributions (see section \ref{two}).
This is the main subject of this paper. This problem appears in all freeze
out calculations up to now we are aware of, and to our knowledge it was not
satisfactorily discussed yet in the literature. It was recognized by some of
those who applied the Cooper-Frye freeze out description before \cite
{Si89,Bu96,BM96}. A possible partial solution was presented in part 2 of ref.%
\cite{Bu96} for noninteracting massless particles, in 1+1 dimension using
the post FO cut J\"uttner ansatz. In \cite{LANL-XXX} it was shown that in an
oversimplified kinetic freeze out model one can obtain the cut J\"uttner
distribution as post FO distribution. In section \ref{three} we complete and
generalize the results of \cite{Bu96} and present an example for the
solution of the freeze out problem. In section \ref{four} we suggest
improvements that go beyond the cut J\"uttner ansatz.

\section{Conservation laws\protect\\across idealized freeze out
discontinuities}

\label{two}

\indent In the zero width limit of the freeze out domain (freeze out
surface), the energy - momentum tensor changes discontinuously across this
surface. Consequently, the four-vector of the flow velocity may also change 
\cite{Cs87,CC94,CG88}. These changes should be discussed in terms of the
conservation laws.

The invariant number of conserved particles (world lines) crossing a surface
element, $d\sigma ^\mu $, is 
\begin{equation}
dN=N^\mu \ d\sigma _\mu \ ,  \label{e00}
\end{equation}
and the total number of all the particles crossing the FO hyper-surface, $S$%
, is 
\begin{equation}
N=\int_SN^\mu \ d\sigma _\mu \ .  \label{efo1}
\end{equation}
If we insert the kinetic definition of $N^\mu $ 
\begin{equation}
N^\mu =\int \frac{d^3p}{p^0}\ p^\mu \ f_{FO}(x,p)\ ,  \label{Ndef}
\end{equation}
into eq. (\ref{e00}) we obtain the Cooper-Frye formula\cite{CF74}: 
\begin{equation}
E\frac{dN}{d^3p}=\int f_{FO}(x,p)\ p^\mu d\sigma _\mu \ ,  \label{e-cf}
\end{equation}
where $f_{FO}(x,p)$ is the post FO phase space distribution of frozen-out
particles which is not known from the fluid dynamical model. The problem is
to choose its form correctly. Usually one assumes that the pre-FO momentum
distribution as well as the post FO distribution are both local thermal
equilibrium distributions, with the same temperature, boosted by the local
collective flow velocity on the actual side of the freeze out surface,
although the post FO distribution need not be a thermal distribution.
Parametrizing the post FO distribution as thermal, $f_{FO}(x,p;T,n,u^\mu )$,
where $T$ does not necessarily coincide with the pre FO temperature, and
knowing the pre-FO baryon current and energy-momentum tensor, $N_0^\mu $ and 
$T_0^{\mu \nu }$, we can calculate the post freeze out quantities $N^\mu $
and $T^{\mu \nu }$ from the relations~\cite{Ta48,Cs87} 
\begin{equation}
\lbrack N^\mu \ d\sigma _\mu ]=0\ \ \ \ {\rm and}\ \ \ \ \ [T^{\mu \nu }\
d\sigma _\mu ]=0,  \label{efo2}
\end{equation}
across a surface element\footnote{%
In numerical calculations the local freeze out surface can be determined
most accurately via self-consistent iteration \cite{Bu96,NL97}.} of normal
vector $d\sigma ^\mu $. Here $[A]\equiv A-A_0$. The post FO distribution is
not a thermal equilibrium distribution, so temperature does not exist,
nevertheless, the conservation laws fix the {\em parameters}, e.g. $T$, $n$, 
$u^\mu $, of our momentum distribution, $f_{FO}(x,p;T,n,u^\nu )$.

To obtain a physically realizable result, in addition we have to check the
condition for entropy increase: 
\begin{equation}
\lbrack S^\mu \ d\sigma _\mu ]\ge 0\ \ \ \ {\rm or}\ \ \ \ \ R=\frac{S^\mu
d\sigma _\mu }{S_0^\mu d\sigma _\mu }\ge 1\,,  \label{efo9}
\end{equation}
where, for both equilibrium and nonequilibrium FO distributions\cite{Cs94} 
\begin{equation}
S^\mu =-\int \frac{d^3p}{p^0}\ p^\mu \ f_{FO}(x,p)\ \left[ \ln \left\{
\left( 2\pi \right) ^3\ f_{FO}(x,p)\right\} -1\right]  \,.
\end{equation}
This condition is not necessary to obtain a solution of the freeze out
problem, but it should always be checked to exclude non-physical solutions.
We have to note at this stage, that the post FO distribution must not be an
equilibrium (or stationary) solution of the Boltzmann Transport Equation,
and consequently on the post FO side the energy flow-, baryon flow- and
entropy flow-velocities may all be different.

We can now remind briefly what the problem of negative contributions to the
Cooper-Frye formula is, and a possible way out. For a FO surface with
time-like normal, both $p^\mu $ and $d\sigma ^\mu $ are time-like vectors,
thus, 
\[
p^\mu d\sigma _\mu >0, 
\]
and the integrand in the integral (\ref{e-cf}) is always positive. For a FO
surface with space-like normal, $p^\mu $ is time-like and $d\sigma ^\mu $ is
space-like, thus, $p^\mu d\sigma _\mu $ can be both positive and negative.
(Note that $p^\mu $ may point now both in the post- and pre- FO directions.)
Thus, the integrand in the integral (\ref{e-cf}) may change sign in the
integration domain, and this indicates that part of the distribution
contributes to a current going back, into the front while another part is
coming out of the front. On the pre-FO side $p^\mu $ is unrestricted and $%
p^\mu d\sigma _\mu $ may have both signs, because we are supposing that
pre-FO phase is in the thermal equilibrium. However, in the zero width limit
of the FO front, it is difficult to understand such a situation. What
happens actually is that internal rescatterings occur inside the finite FO
domain and feed particles back to the pre-FO side to maintain the thermal
equilibrium there. On the post-FO side, however, we do not allow
rescattering and back scattering any more. If a particle has passed the
freeze out domain it cannot scatter back. In other words, the post-FO
distribution should have the form\cite{Si89,Bu96}, 
\begin{equation}
f_{FO}^{*}(x,p,d\sigma ^\mu )=f_{FO}(x,p)\Theta (p^\mu \ d\sigma _\mu )\,,
\label{cut}
\end{equation}
where $\Theta(x)$ is the step function. Consequently, this distribution
cannot be an ideal gas distribution. (On the pre-FO side, the distribution
may or not be ideal). The conservation laws across a small element of the
freeze out front with space-like normal take the form: 
\begin{equation}
\int_S\left( \int \frac{d^3p}{p^0}\ \ f_{FO}^{*}(x,p,d\sigma ^\gamma )\
p^\mu \right) \ d\sigma _\mu =\int_SN_0^\mu (x)\ d\sigma _\mu \ ,
\label{efo7}
\end{equation}
\begin{equation}
\int_S\left( \int \frac{d^3p}{p^0}\ \ f_{FO}^{*}(x,p,d\sigma ^\gamma )\
p^\mu p^\nu \right) \ d\sigma _\mu =\int_S\ T_0^{\mu \nu }(x)\ d\sigma _\mu
\ .  \label{efo8}
\end{equation}

\section{The allowed momentum region for space-like FO}

\label{two-add}

Let us assume that the FO process happens in the positive x direction, in
other words we go in the positive $x$-direction from the pre FO domain to
the post FO domain. The FO hyper-surface has a space-like normal ($d\sigma
^\mu d\sigma _\mu =-(d\sigma )^2$), so that $d\sigma _\mu $ is orthogonal
to the hyper-surface (i.e. to the time-like tangent vector of the surface $%
t^\mu $, $t^\mu d\sigma _\mu =0$) and points into the post FO (positive $x$-
) direction (while $d\sigma ^\mu $ points in the pre FO direction).

Depending on the reference frame, the space-like FO front can propagate both
in the positive or negative $x$-direction, or it can be Lorentz transformed
into its own rest frame, the Rest Frame of the Front (RFF), where $d\sigma
_\mu $ $ =(0,1,0,0)d\sigma $. \ \  In other reference frames $d\sigma _\mu $ 
$ =\gamma
(v,1,0,0)d\sigma $, where $\gamma = $ $1/\sqrt{1-v^2}$. The parameter $%
v=d\sigma _0/d\sigma _x$ $=-t^x/t^0$ is frame dependent and may be both
positive and negative.

In order to use the parameter $v$ in the following discussion, we have to
select a given reference frame and fix the value in that frame. Let us
choose the frame {\em comoving with the peak} of the post FO invariant
momentum distribution. Following ref. \cite{Bu96} let us denote this frame
as the Rest Frame of the Gas (RFG). 
Note, however, that, contrary to what its name seems to suggest, this
frame is {\em not the Local Rest frame} of the post FO matter, since the
post FO distribution is not spherically symmetric in momentum space! Thus,
for the following discussion we {\em define} $v$ in the 
RFG frame as 
\[
v\equiv \left. \frac{d\sigma _0}{d\sigma _x}\right| _{RFG}\,,
\]
or we can also define and have the same $v$-value, by using the velocity of
the peak of the post FO distribution, $u_{RFG}^\mu $ in the RFF. (Note: this
is not equal with the post FO flow velocities, neither with the Eckart-, nor
with the Landau- flow velocity!) Thus, 
\[
v=\left. \frac{u_{RFG}^x}{u_{RFG}^0}\right| _{RFF}\,,
\]
and this means that in the RFF the peak four-velocity has the form $%
u_{RFG}^\mu =\gamma (1,v,0,0)|_{RFF}\,.$ This velocity $v$ can also be both
positive and negative (!) in RFF, i.e. the peak velocity may point to the
post FO direction (as we would expect), and also in the pre FO direction in
RFF which seems to be confusing. This is, however, not a problem in itself
because both flow velocities (Eckart and Landau) are always positive on the
post FO side. As we will see later, in special cases it is possible that we
obtain negative post FO peak velocity for positive pre FO flow velocities.
This indicates we have to discuss the importance of the cut by $\Theta
(p^\mu d\sigma _\mu )$, otherwise one might be tempted to believe that this
cut is not affecting the post FO distribution too much, and the correct
treatment causes only a few percent cut which is negligible. We show in the
following that
this is not the case.

\begin{figure}[tbh]
\insertplot{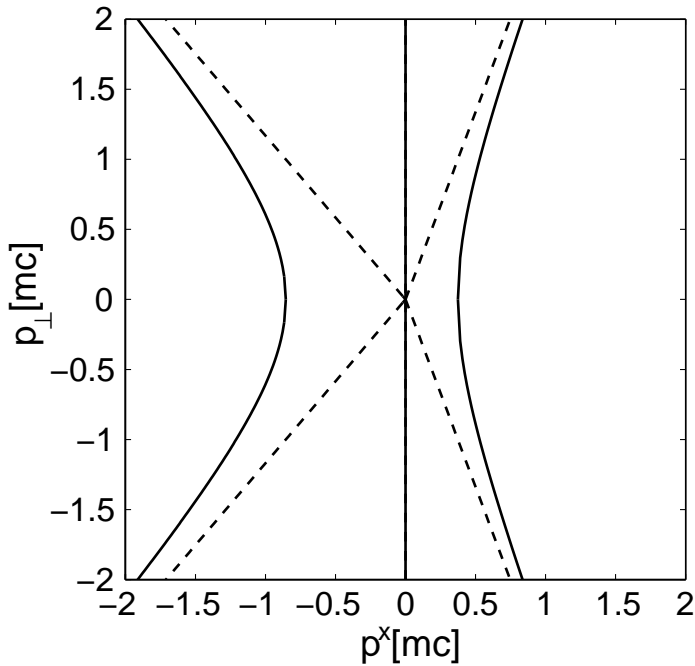}
\vspace{0.5cm}
\caption[]{\protect\footnotesize 
The boundaries of the cut post FO
distributions in the phase space indicating the regions where particles are
allowed to freeze out. These boundaries are hyperboloids, for RFF velocities
of: $v=0.65$, $v=0$ and $v=-0.35$ (from left to right). Particles to the
right of these hyperboloids may freeze out. While for large positive $v$
values the cut is a small perturbation, for moderate or particularly for
negative values of $v$ the cut is far from negligible. The dotted lines
show the massless limits. }
\label{f-hyp}
\end{figure}

The $p^\mu d\sigma _\mu >0$ requirement in the 
RFG frame means that only momenta with component 
\begin{equation}
p^x\ge -v\sqrt{m^2+(p^x)^2+(p_{\perp })^2}\,,  \label{p-cut}
\end{equation}
contribute to the post FO momentum distribution. This means that the
boundaries of the allowed domain in the $[p^x,p_{\perp }]$- plane are
hyperbolas in the post FO RFG 
\begin{equation}
\frac{(p^x)^2}{\gamma ^2-1}-(p_{\perp })^2=m^2\,,  \label{hyperbola}
\end{equation}
and the domains of the positive $x$- side of the corresponding hyperboloids
in the 3-dimensional momentum space may contribute to the FO distribution
(Fig. \ref{f-hyp}). In the massless limit the hyperboloids become cones
around the $x$-axis and centered at the origin.

\bigskip

\bigskip

\paragraph{Dominant case.}

\ For $v=0$ the hyperboloid becomes a plane boundary at $p^x\ge 0$ in RFG,
i.e. the boundary cuts the J\"{u}ttner distribution in the middle (because
in the RFG the peak of the distribution is centered at $\vec{p}=0$). If $%
v>0 $, the peak velocity points to the post FO direction and the
cut-hyperboloid is fully in the negative $p^x$ region, thus for large $v$
values only a smaller fraction is excluded from the full J\"{u}ttner
distribution. Most of the particles described by a full J\"{u}ttner
distribution freeze out in this case. We will consider this as the dominant
FO case.

On the other hand, as we can see in Fig. \ref{f-hyp}, for the $v < 0$ case
the peak flow points back into the pre FO direction, and only a small
fraction of a full J\"uttner distribution will freeze out. The cut
eliminates the major part of the particles. This situation leads to less
frozen out particles, but it yields an unusual post FO distribution.

\section{Conserved currents for cut J\"uttner distribution}

\label{three}

We now study the particular case where $f_{FO}$ is a J\"{u}ttner (or
relativistic Boltzmann \cite{juttner,Cs94}) distribution and so $f_{FO}^{*}$
is a cut J\"{u}ttner distribution: 
\begin{equation}
f_{FO}^{*}(p)=f^{Juttner}(p)\Theta (p^\mu d\sigma _\mu )=\frac{%
\Theta (p^\mu d\sigma _\mu )}{(2\pi \hbar )^3}\exp \left( \frac{\mu -p_\mu
u_{RFG}^\mu }T\right) \,,
\end{equation}
where $\mu $ is the chemical potential related to the invariant scalar
density $\hat{n}$, of the non-cut J\"{u}ttner distribution as $\mu =$ $T\ln
\left[ \hat{n}(2\pi \hbar )^3/\left( 4\pi m^2TK_2\left( {\frac mT}\right)
\right) \right] ,$ and $u_{RFG}^\mu $, $\hat{n}$ and $T$ are {\em parameters%
} of the distribution $f_{FO}^{*}$ originating from the full J\"{u}ttner
distribution. These are not the flow velocity, proper density and
temperature of the cut J\"{u}ttner distribution! The cut J\"{u}ttner
distribution is not a thermal equilibrium distribution, e.g. it does not
have a temperature at all.

This distribution for massless particles was considered in part 2 of ref.%
\cite{Bu96} (also in~\cite{Si89}). The cut selects particles with momenta $%
p^\mu d\sigma _\mu >0$.

We can evaluate the baryon four-current, $N^\mu$, by inserting the cut
J\"uttner distribution into the definition, eq. (\ref{Ndef}), and we get a
time directed-, $N^0$, as well as a spatial component, $N^x$, (where $x$ is
the direction of the spatial component of the FO-normal, $d\sigma^\mu$, in
RFG). In ref.\cite{Bu96} the spatial component, $N^x$, of the four current
was not evaluated, so seeing only the zeroth component, $N^0$, the
unsuspecting reader might have believed falsly, that RFG (Rest Frame of the
Gas) is the Local Rest frame of the gas. Performing the calculation, in the
post FO RFG frame the baryon current reads as 
\begin{equation}
\begin{array}{rl}
N^0= & \frac{\tilde{n}}{4} \left\{ vA + a^2 j \left[ (1{+}j) K_2(a) - {\cal K%
}_2(a,b) \right] + \right. \\ 
& \left . \hspace*{1cm} j \frac{b^3 v^3}{3} e^{-b} \right\} \hspace*{1cm} 
\stackrel{\scriptscriptstyle m=0}{\longrightarrow }\ \,\tilde{n}(\mu ,T)%
\frac{v+1}{2}\,, \\ 
N^x= & \frac{\tilde{n}}{8}\left[ \left(1-v^2\right) A-a^2e^{-b}\right] 
\stackrel{\scriptscriptstyle m=0}{\longrightarrow }\ \,\tilde{n}(\mu ,T)%
\frac{1-v^2}{4}\,,
\end{array}
\label{NcJ}
\end{equation}

\noindent
where $j={\rm sign}(v)$, \ $\tilde{n}=8\pi T^3e^{\mu /T}(2\pi \hbar )^{-3}$,
\ $a=\frac mT$, \ so that $\hat{n}(\mu ,T)=\tilde{n}a^2K_2(a)/2$ is the
invariant scalar density of the symmetric J\"{u}ttner gas, $b=a/\sqrt{1-v^2}$%
, \ $v=d\sigma _0/d\sigma _x$, \ $A=(2+2b+b^2)e^{-b}$, \ and 
\[
{\cal K}_n(z,w)\equiv \frac{2^n(n)!}{(2n)!}\ z^{-n}\!\!\int_w^\infty
\!\!\!dx\ (x^2{-}z^2)^{n{-}1/2}\ \ e^{-x}\ , 
\]
i.e. ${\cal K}_n(z,z)=K_n(z)$. \ \ \ \ Just as in case of the non-cut
distributions the cut J\"{u}ttner distribution yields few modified Bessel
functions in the expression of the four currents, while the relativistic
Fermi and relativistic Bose distributions lead to a series of these
functions. When evaluating the limits\ \ we used the relation \ \ ${\cal K}%
_n(a,b)\stackrel{\scriptstyle a=b}{\longrightarrow }$ $K_n(a)$ $\stackrel{%
\scriptstyle a=0}{\longrightarrow }$ $2^{n-1}(n-1)!a^{-n}$. This baryon
current may then be Lorentz transformed into the Eckart Local Rest (ELR)
frame of the post FO matter, which moves with $u_E^\mu =N^\mu /(N^\nu N_\nu
)^{1/2}=\gamma _E(1,v_E,0,0)|_{RFG}$ in the RFG,\ \ or alternatively into
the Rest Frame of the Freeze out front (RFF), where $d\sigma _\mu
=(0,1,0,0)(d\sigma )^2|_{RFF}$ and the velocity of the RFG is $%
u_{E,RFG}^\mu =\gamma _\sigma (1,v,0,0)|_{RFF}$. Then the Eckart flow
velocity of the matter represented by the cut J\"{u}ttner distribution
viewed from the RFF is $u_E^\mu =\gamma _c(1,v_c,0,0)|_{RFF}$, where $%
v_c=(v+v_E)/(1+vv_E)$.

The proper density (i.e. the density in the ELR frame) is obtained as 
\begin{equation}
n(\mu,T,v) = \sqrt{N^\nu N_\nu} \,.  \label{pro-n}
\end{equation}
Note that the proper density of the cut J\"uttner distribution, $n$, is
reduced compared to the proper density of the complete spherical J\"uttner
distribution, $\hat{n}$.

The energy momentum tensor in the post FO RFG is 
\begin{equation}
\begin{array}{rl}
T^{00}= & \frac{3\tilde{n}T}2\left\{ \frac{ja^2}2\left\{ (1{+}j)\left[ K_2(a)%
{+}\frac a3K_1(a)\right] {-}\left[ {\cal K}_2(a,b){+}\frac a3{\cal K}%
_1(a,b)\right] \right\} {+}Bv\right\} \!, \\ 
T^{0x}= & \frac{3\tilde{n}T}4\left\{ (1-v^2)B-{\frac{a^2}6}%
(b+1)e^{-b}\right\} , \\ 
T^{xx}= & \frac{\phantom{1}\tilde{n}T}2\left\{ j\frac{a^2}2\left[ (1{+}%
j)K_2(a)-{\cal K}_2(a,b)\right] +v^3B\right\} , \\ 
T^{yy}= & \frac{3\tilde{n}T}4\left\{ v(1{-}\frac{v^2}3)B+{\frac{ja^2}3}%
\left[ (1+j)K_2(a)-{\cal K}_2(a,b)\right] -{\frac{{va^2}}6}(b{+}%
1)e^{-b}\right\} ,
\end{array}
\label{TcJ}
\end{equation}
where $B=(1+b+b^2/2+b^3/6)e^{-b}$ and $T^{zz}=T^{yy}$. This energy-momentum
tensor may then be Lorentz transformed into the Landau Local Rest (LLR)
frame of the post FO matter, which moves with $u_L^\mu $ in the RFG, or into
the Rest Frame of the Freeze out Front (RFF) where $d\sigma ^\mu
=(0,1,0,0)d\sigma $. Alternatively both can be transformed to the frame
where we want to evaluate the conservation laws, eq. (\ref{efo2}), and the
parameters of the post FO, cut J\"{u}ttner distribution can be determined
so, that it satisfies the conservation laws. In the massless limit the
energy momentum tensor in the RFG is: 
\[
T^{00}=3\tilde{n}T\left( v{+}1\right) /2\,,\ \ \ T^{0x}=3\tilde{n}T\left( 1{-%
}v^2\right) /4\,,\ \ \ T^{xx}=\tilde{n}T\left( v^3{+}1\right) /2 
\]
and $T^{zz}=T^{yy}=(T^{00}-T^{xx})/2$.

In addition we have to check the entropy condition. In the 
RFG frame the entropy current reads 
\begin{equation}
\begin{array}{rl}
S^0= & \frac{\tilde{n}}4\left\{ (1{-}\frac \mu T)vA+6vB+(1-\frac \mu
T)a^2j\left[ (1{+}j)K_2(a)-{\cal K}_2(a,b)\right] +\right. \\ 
& \left. ja^2\left[ (1{+}j)K_1(a)-{\cal K}_1(a,b)\right] \right\} %
\hspace*{0.5cm}\stackrel{\scriptscriptstyle m=0}{\longrightarrow }\ \,%
\hspace*{0.5cm}\frac{\tilde{n}(\mu ,T)}2(v+1)\left( 4-\frac \mu T\right) \,,
\\ 
S^x= & \frac{\tilde{n}}8\left[ \left( 1-v^2\right) \left( 1-\frac \mu
T\right) A+6\left( 1-v^2\right) B-a^2\left( 2+b-\frac \mu T\right)
e^{-b}\right] \\ 
& \hspace*{5.7cm}\stackrel{\scriptscriptstyle m=0}{\longrightarrow }\ \,%
\frac{\tilde{n}(\mu ,T)}4(1-v^2)\left( 4-\frac \mu T\right) \,.
\end{array}
\label{ScJ}
\end{equation}
Note that in the $m=0$ limit the vectors $S^\mu $ and $N^\mu $ are parallel
to each other. This is explained by Figure \ref{f-hyp}, which shows that in
the RFG the cut in the $m=0$ limit becomes a central cone, and since the
distribution is centrally symmetric in this frame, the integrals will be
proportional to each other.

\subsection{Solubility of the freeze out problem}

The situation is non-trivial and we have to take into account the possible
directions of the flow and of $d\sigma_\mu$. Note: we must not assume that
the flow is parallel to the freeze out direction.

Let us start on the pre FO side labeled by `0'. Here in the LR frame $%
u_0^\mu $ $=$ $(1,0,0,0)|_{pre-LR}$ and we can choose the $x$-direction in
this frame to point into the FO direction, so that $d\sigma _\mu $ $=$ $%
\gamma _0(v_0,1,0,0)d\sigma |_{pre-LR}$. We assume that we know the FO
hypersurface, i.e. we know $v_0$. Then, in this frame the conservation laws
have three nonvanishing components yielding three known parameters $N_0^\mu
d\sigma _\mu $, $T_0^{0\mu }d\sigma _\mu $, and $T_0^{x\mu }d\sigma _\mu $.

To find the solution we need these values in the 
RFG frame. However, the 3-dimensional direction of the $x$-axis will not
change because the front is assumed to be isotropic in its own $[y,z]$%
-plane. Thus, in the  
RFG the peak flow parameter is $u_{RFG}^\mu =(1,0,0,0)|_{RFG}$, and the
normal of the FO front is $n_\mu =\gamma (v,1,0,0)$. Note that $v\ne v_0$!
Furthermore, let us recall that the parameter $v$ determines the post FO
peak flow parameter in RFF, $u_{RFG}^\mu =\gamma (1,v,0,0)|_{RFF}$ (where $%
d\sigma _\mu $ $=$ $(0,1,0,0)d\sigma |_{RFF}$).

Consequently the conservation laws (\ref{efo7},\ref{efo8}) yield three
nonvanishing equations in the 
RFG frame, 
\[
\lbrack N^\mu d\sigma _\mu ]=0\,,\ \ [T^{0\mu }d\sigma _\mu ]=0\ \ {\rm and}%
\ \ [T^{x\mu }d\sigma _\mu ]=0\,, 
\]
which can determine the three unknown {\em parameters} of the post FO
cut-J\"{u}ttner distribution, $v$, $T$, and $n$ (or $\mu $). While the first
equation is an invariant scalar, the remaining two are components of a
4-vector, so they should be transformed into the same reference frame, i.e.,
to RFG. Since we evaluated the quantities based on the cut-J\"{u}ttner
distribution in the RFG, we also need the pre FO quantities in the 
RFG. These can be determined by using the standard fluid dynamical form of $%
T^{\mu \nu }$ as seen from the RFG. From this frame the pre FO flow velocity
is given by the difference of the pre- and post-FO flow velocities: $u_0^\mu
=\gamma _{0R}(1,v_{0R},0,0)|_{RFG}$, where $v_{0R}=(v_0-v)/(1-v_0\,v)$.

\begin{figure}[tbh]
\insertplot{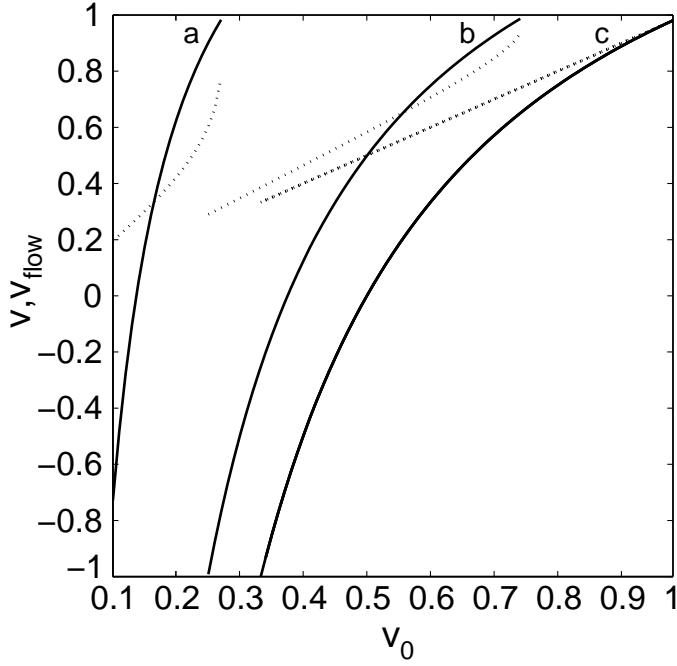}
\vspace{0.5cm}
\caption[]{\protect\footnotesize 
Change of velocities in freeze out of
QGP to hadronic matter described by massless cut J\"{u}ttner distribution.
The final velocity {\em parameters} (full lines) of the cut J\"{u}ttner
distribition are plotted versus the initial flow velocity of QGP measured in
the Rest Frame of the Front (RFF) for cases: (a) $n_0=$1.2 fm$^{-3}$, $%
T_0=60 $ MeV, $\Lambda _B=225$ MeV, (b) $n_0=$0.1 fm$^{-3}$, $T_0=60$ MeV, $%
\Lambda _B=80$ MeV (c) $n_0=$1.2 fm$^{-3}$, $T_0=60$ MeV, $\Lambda _B=0$
MeV. Observe that for small initial flow velocities the center of the cut
J\"{u}ttner distribution moves backwards, although all the particles which
are allowed to freeze out move forward. Thus, the post FO {\em baryon flow
velocities} (dotted lines) are positive. Note the large acceleration caused
by the released latent heat in cases (a) and (b). $\Lambda _B=B^{1/4}$,
where $B$ is the bag constant.}
\label{f-v-v0}
\end{figure}

In the general case the solution can be obtained numerically. In the $m=0$
limit the solution is simpler and gives an interesting insight into the
problem. The continuity equation leads to the equation 
\begin{equation}
Q_1^{-2} (v+1)^3 + v - 1 = 0\,,  \label{vcont}
\end{equation}
where $Q^{-1} = Q_1^{-1}(\mu,T) = \tilde{n}(\mu,T)/(4 n_0 \gamma_0 v_0 )\,, $
which leads to a 3rd order equation and can be solved for $v$ alnalytically.
The energy equation, $[T^{0\mu} d\sigma_\mu]=0 $, leads to the same equation
but with another coefficient $Q^{-1} = Q_2^{-1}(\mu,T) = 3 T \tilde{n}%
(\mu,T)/(4 e_0 \gamma_0 v_0 )\,, $ thus, these two equations can have one
and the same solution for $v$, only if the two coefficients, $Q_1$ and $Q_2$%
, are equal, which results in 
\[
T = \frac{1}{3} \frac{e_0}{n_0} \,, 
\]
and the solutions of both 3rd order equations yield the same expression: 
\begin{equation}
v = v_{3rd} (\mu) = Q^{2/3} \left\{ \left[ 1{+}\sqrt{1{+}Q^2/27}
\right]^{1/3} + \left[ 1{-}\sqrt{1{+}Q^2/27} \right]^{1/3} \right\} - 1 \,.
\label{v3rd}
\end{equation}
Then, dividing the equation $[T^{0\mu} d\sigma_\mu]=0 $, by the equation $%
[T^{x\mu} d\sigma_\mu]=0 $, yields another 3rd order equation for $v$: 
\[
R_0 v^3 + 3 v^2 + 3(2-R_0)v + 3-2R_0 =0 \,, 
\]
where $R_0=e_0 v_0 / p_0$. This equation can be solved analytically and
yields one physical root, $v=2-3/R_0$ (and two unphysical ones $v=-1$).
Inserting $v$ then into eq. (\ref{vcont}), we can obtain the resulting
chemical potential, $\mu$, also.

\begin{figure}[htb]
\insertplot{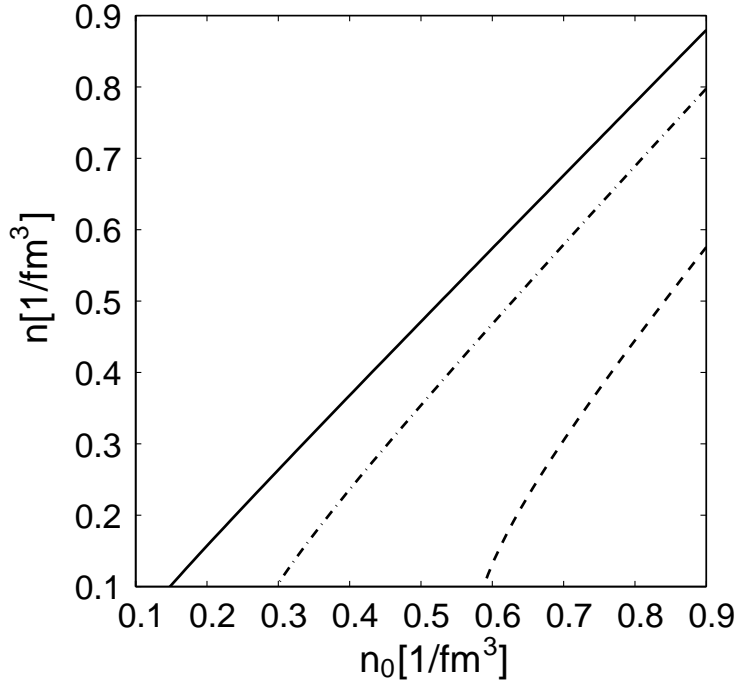}
\vspace{0.5cm}
\caption[]{\protect\footnotesize 
The final baryon density, $n$, as a function
of the pre FO baryon density. The baryon density decreases in the freeze out
process for cases $v_0=0.5$, $T_0=50$ MeV, (a) $\Lambda_B = 80$ MeV (full
line), (b) $\Lambda_B =120$ MeV (dashed-dotted line) and (c) $\Lambda_B =160$
MeV (dashed line).}
\label{f-n-n0}
\end{figure}

The possibility of this simple analytic solution is a consequence of the
fact that in the $m=0$ limit the cut of the J\"{u}ttner distribution is made
along central cones in the RFG, which then divide the energy and the baryon
charge exactly in the same proportions.

As an illustration we studied the freeze out of quark-gluon plasma (QGP) to
cut J\"uttner gas, in the massless limit. The pre FO side QGP is described
by the most simple bag-model EoS (eqs. (5.28-33) in ref. \cite{Cs94}), thus
local equilibrium is assumed and all pre FO parameters are assumed to be
known including the baryon, energy-momentum and entropy currents.

On the post FO side these currents were evaluated earlier in this section,
and the equations arising from the conservation laws, (\ref{efo2}), were
solved as presented above. Figure \ref{f-v-v0} indicates the change of flow
velocity during freeze out. Physical solution exists only for positive
initial velocities, $v_0 \ge 0$. The velocity {\em parameter} (!) of the
post FO cut J\"uttner distribution varies from $-1$ to $+1$, but the post FO
Eckart flow velocity is of course always positive in RFF. Thus, the post FO
baryon current is also positive in RFF (this is obvious since we do not
allow any particle to cross the front backwards), and consequently, the pre
FO current and $v_0$ should also be positive because of the continuity
equation. For small initial velocities, $v_0 \longrightarrow 0 $, the post
FO velocities approach zero also, but for moderate velocities, deduced
recently from experiments, $v = 0.3 - 0.7$, the difference between the post
and pre FO flow velocities may be essential.

In order to show the effect of these modifications compared to the original
Cooper-Frye treatment (where the increase of the flow velocity is ignored)
we can consider case (a) in Figure 2. The cut J\"uttner distribution always
leads to an exponential $p_t$ spectrum, but according the new modified
treatment starting from $v_0 = 0.2$c the post FO flow velocity increases to $%
v_{flow}=0.4$c, while the post FO parameter velocity (which determines the $%
p_t$ spectrum) increases to $v=0.6$c. This corresponds to an increase of the
slope parameter, $T_{slope}$, by 60\%! This is due to the large latent heat
arising from the large value of the bag constant taken in case (a). In case
(b) the the same effect is present but it is weaker. This change of the flow
velocity is a basic feature of the correct freeze-out treatment, and it is a
consequence of the conservation laws and not of the positivity requirement
of $p^\mu d\sigma_\mu$ in spacelike FO. Thus, the flow velocity change
occurs both in spacelike and in timelike freeze-out. This effect can cause
for example the conversion of latent heat to collective kinetic energy and
not to heat if the freeze-out coincides with an exotherm phase transition 
\cite{CC94}.

Figure \ref{f-n-n0} shows that the baryon density, (\ref{pro-n}), decreases
in the freeze out process. This is connected to the fact that the post FO
flow velocities are above the pre FO ones, as shown in Figure \ref{f-v-v0}.

We should mention that the post FO temperature {\em parameter} of the cut
J\"uttner distribution becomes rather high, about an order of magnitude
higher than the pre FO temperature. However, we have to recall that the term
temperature is not applicable for a non-equilibrium distribution, therefore
this result has no physical significance, it just illustrates the
parametrization of the distribution of the assumed cut J\"uttner shape.

\begin{figure}[htb]
\insertplot{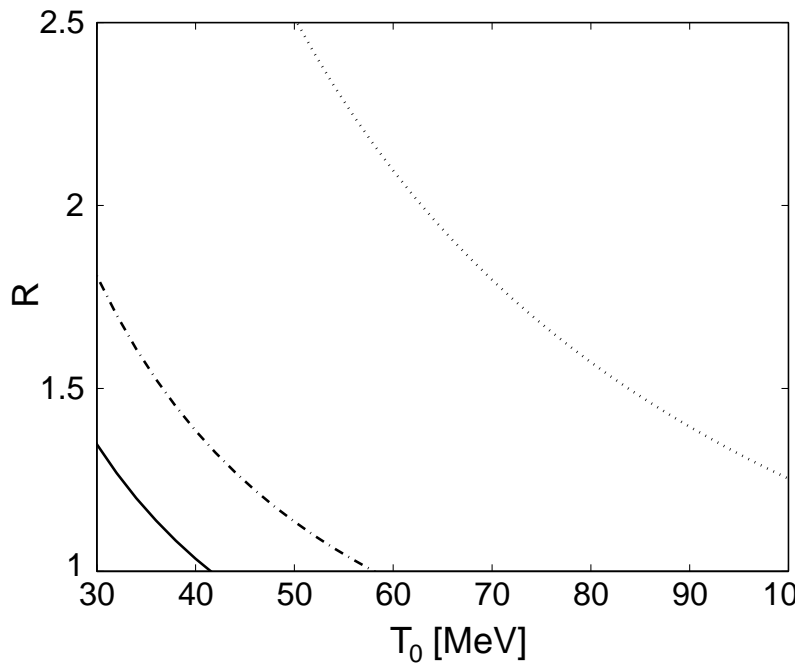}
\vspace{0.5cm}
\caption[]{\protect\footnotesize The ratio of post FO and pre FO entropy
currents transverse to the freeze out front. Freeze out can be physically
realized if $R>1$. The entropy condition is tested for three cases: (a) $%
n_0= $ 0.1 fm$^{-3}$ $v_0=0.5$, $\Lambda_B = 80$ MeV (full line), (b) $n_0=$
0.5 fm$^{-3}$ $v_0=0.5$, $\Lambda_B = 80$ MeV (dashed line) and (c) $n_0=$
1.2 fm$^{-3}$ $v_0=0.5$, $\Lambda_B =225$ MeV (dotted line). }
\label{f-R-t0}
\end{figure}

Finally we have to check the entropy condition for these solutions. As we
know \cite{CC94,DR87,CK92} QGP can freeze out to hadronic matter with
entropy production only if the QGP is supercooled or considerably
supercooled. This remains valid for the cut J\"uttner assumption as post FO
distribution also. With most parametrizations only low temperature QGP is
able to freeze out. For the cut J\"uttner gas we cannot speak of a critical
temperature, because this gas is not in equilibrium and consequently cannot
be in phase equilibrium either. Still this distribution can be attributed an
entropy current by its kinetic definition, and the entropy condition can be
checked (Fig. \ref{f-R-t0}).

In reality the entropy condition is not so stringent as Figure \ref{f-R-t0}
indicates. In this illustrative study the post FO EoS had relatively few
degrees of freedom to accomodate the high entropy content of QGP. By
including many post FO mesons and other hadronic degrees of freedom in our
post FO EoS, the entropy condition can be satisfied in a much wider range of
parameters.

\section{Freeze out distribution from kinetic theory}

\label{four}

We have seen that taking the cut J\"uttner distribution as an ansatz for the
post FO distribution, we can solve the freeze out problem formally. Although
we can satisfy all requirements, the obtained parameter values make it
questionable whether the cut J\"uttner ansatz is an adequate assumption. The
shape of the distribution with the sharp cuts is also a rather unphysical
feature of the distribution.

To obtain a more realistic, and physically better applicable FO
distributions, we should evaluate the distribution in more physical,
microscopic non-equilibrium models. Kinetic theory is a straightforward
candidate for this task.

A first very simplified attempt to solve the freeze out problem dynamically
in one dimensional kinetic model \cite{LANL-XXX} returned the cut J\"uttner
distribution also, but only in highly unrealistic situations: only when the
model yielded incomplete freeze out. Thus, further work is needed to find
physically realistic post freeze out distributions in kinetic models or in
other dynamical microscopic models.

\section{Conclusions}

The importance of taking into account conservation laws in the description
of the freeze out process is pointed out. For freeze out across
hypersurfaces with space-like normals the approach suggested by Bugaev,\cite
{Bu96} assuming cut J\"uttner distribution as post freeze out distribution
is worked out, and the freeze out problem was solved as an example for QGP
freezing out into a cut J\"uttner gas. This calculation indicates that
results including the Cooper-Frye freeze out procedure should be
reconsidered and new emphasis should be given to the precise evaluation of
the post freeze out particle distributions.

The deviation from the earlier Cooper-Fry approach (where changes of flow
velocity, density and temperature were ignored) is apparent if the pre FO
matter has large energy content in form of compressional energy, latent
heat, or in any other way, which is not present in the post FO,
noninteracting matter. As this post FO matter is not necessarily in thermal
equilibrium, we cannot consider it as a thermal phase with equilibrium
thermodynamical parameters. Thus, this idealized approach assuming a FO
surface is always assuming a discontinuity irrespective of what was the
phase of the pre FO matter. Nevertheless, this treatment leads to the
strongest modifications in cases when a first order phase transition with
large latent heat is coupled to the freeze out process.

Here we have considered an idealized transition as a discontinuity across a
hypersurface. In as much as the flow across the surface is stationary our
results are valid irrespective of the surface thickness, because we used
only conservation laws. On the other hand in heavy ion reactions the flow
across the surface can be considered stationary only if it is 1-2 fm wide.
With purely kinetic freeze out this is not a very realistic assumption.\cite
{BB95} On the other hand rapid hadronization from supercooled QGP may
satisfy the required conditions and the sharp surface approximation is then
realistic.\cite{CM95}

\section*{Acknowledgement}

This work is supported in part by the Research Council of Norway, PRONEX
(contract no. 41.96.0886.00), FAPESP (contract no. 98/2249-4) and CNPq.
Cs. Anderlik, L.P. Csernai and Zs.I. L\'{a}z\'{a}r are thankful for the
hospitality extended to them by the Institute for Theoretical Physics of the
University of Frankfurt where part of this work was done. L.P. Csernai is
grateful for the Research Prize received from the Alexander von Humboldt
Foundation.

\end{document}